\documentclass[10pt,twocolumn,letterpaper]{article}

\usepackage[pagenumbers]{iccv}      

\definecolor{iccvblue}{rgb}{0.21,0.49,0.74}
\usepackage[pagebackref,breaklinks,colorlinks,allcolors=iccvblue]{hyperref}

\def\paperID{*****} 
\def\confName{ICCV}
\def\confYear{2025}

\title{Provenance Analysis of Archaeological Artifacts via Multimodal RAG Systems}

\author{
\textbf{Tuo Zhang\thanks{These authors contributed equally to this work.}  \textsuperscript{1}, 
Yuechun Sun\textsuperscript{\thefootnote}\textsuperscript{2}, 
and Ruiliang Liu\textsuperscript{3}}\\[6pt]
\textsuperscript{1}Museus\\
\textsuperscript{2}University of Science and Technology of China\\
\textsuperscript{3}British Museum
}

\begin{document}
\maketitle
\begin{abstract}
In this work, we present a retrieval-augmented generation (RAG)-based system for provenance analysis of archaeological artifacts, designed to support expert reasoning by integrating multimodal retrieval and large vision-language models (VLMs). The system constructs a dual-modal knowledge base from reference texts and images, enabling raw visual, edge-enhanced, and semantic retrieval to identify stylistically similar objects. Retrieved candidates are synthesized by the VLM to generate structured inferences, including chronological, geographical, and cultural attributions, alongside interpretive justifications. We evaluate the system on a set of Eastern Eurasian Bronze Age artifacts from the British Museum. Expert evaluation demonstrates that the system produces meaningful and interpretable outputs, offering scholars concrete starting points for analysis and significantly alleviating the cognitive burden of navigating vast comparative corpora.
\end{abstract}    
\section{Introduction}
\label{sec:intro}

Within archaeology, art history, and museology, typological analysis is a foundational methodology for understanding ancient material culture~\cite{sorensen1997material}. Its epistemological basis parallels other scientific methods: the systematic comparison of unknown objects against established reference materials to extrapolate chronological and cultural information~\cite{smith2012approaches}. For example, scholars may date a Chinese Bronze Age ritual vessel by identifying stylistic parallels with securely dated artifacts, whether through inscriptional evidence or stratified archaeological contexts~\cite{luo2025high}. 

While typological analysis draws on diverse theoretical frameworks, from Social Darwinism~\cite{o2000darwinian} to processual archaeology~\cite{lyman2004history}, its core logic remains straightforward. However, the method faces increasing challenges in academic and museum practice. It relies heavily on individual expertise and tacit knowledge, introducing variability and raising concerns about transparency and objectivity~\cite{currie2022speculation}. Researchers also operate under uneven access conditions, from reliance on published catalogs to privileged examination of objects in storage, fundamentally shaping analytical outcomes~\cite{dean2025museum}. Further, each artifact presents a near-infinite array of stylistic features, requiring selective prioritization often guided by Bayesian-like reasoning shaped by disciplinary training~\cite{otarola2022beyond}. 

To mitigate these methodological challenges, this study proposes a retrieval-augmented generation (RAG)-based system for provenance analysis, designed to systematically replicate and enhance expert reasoning processes. In collaboration with domain specialists, we have selected a collection of bronze artifacts exhibiting Eastern Eurasian steppe stylistic traditions from the Bronze Age and Early Iron Age periods (ca. 1500–200 BCE) as our target dataset, which are currently unpublished and housed within the early China section of the British Museum. Evaluation results indicate that the system can generate insightful and suggestive outputs that provide researchers with concrete starting points for analysis, effectively narrowing the search space and alleviating the cognitive burdens of navigating heterogeneous comparative materials.



\section{System Design}
\label{sec:method}

\begin{figure*}[t]
 \centering
 \vspace{-5mm}
 \includegraphics[width = \linewidth]{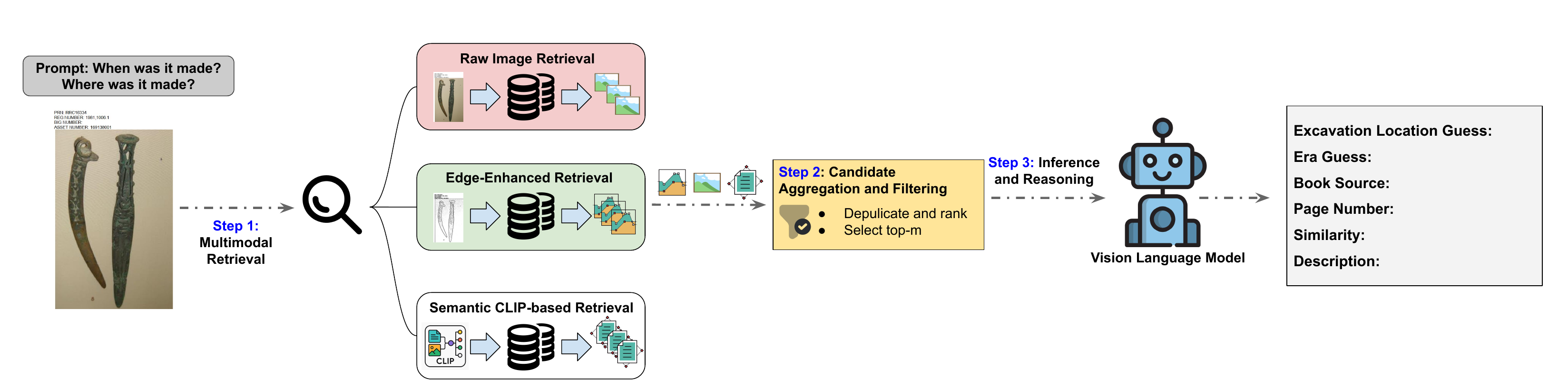}
    \caption{The illustration of the proposed system. the target image is processed through (1) multimodal retrieval (raw, edge-enhanced, and CLIP-based), (2) candidate aggregation and filtering, and (3) inference and reasoning with a vision-language model to predict provenance, era, supporting references, and explanations.}
    \label{fig:system}
    \vspace{-2mm}
\end{figure*}

As shown in the Figure~\ref{fig:system}, our system consists of three core components: \textbf{External Knowledge Construction}, \textbf{Multimodal Retrieval}, and \textbf{Inference and Reasoning}.

\subsection{External Knowledge Construction}
The external knowledge base is primarily composed of archaeological monographs, catalogs, and excavation reports. Source material in PDF format are converted into structure Markdown files to enable efficient text retrieval. However, since many PDFs are low-quality scans rather than digitized texts, text extraction and search can be unreliable. To address this, we additionally extract and index figures, plates, and visual layouts, constructing an image repository to enable complementary image-based querying.

\subsection{Multimodal Retrieval}

To fully leverage both the textual and visual modalities in archaeological documents, we design a three-fold retrieval module. These three retrieval strategies run in parallel and serve as structural complements to each other.

\noindent \textbf{Strategy 1: Raw Image Retrieval.} The input query image is directly compared against all reference images in the image-only candidate database using cosine similarity. The top-$k$ visually similar images are selected, and the corresponding contextual paragraphs in their source documents are retrieved as candidate reference knowledge.

\noindent \textbf{Strategy 2: Edge-Enhanced Retrieval via Gaussian Filtering.} To address the frequent presence of line drawings and archaeological sketches in the external database, we apply Gaussian-filter-based edge detection on query images to enhance the recognition. The filtered image highlights structural edges and contours, which are then used to query the database via cosine similarity. The Gaussian filter is defined as:

\[
G(x, y) = \frac{1}{2\pi \sigma^2} \exp\left(-\frac{x^2 + y^2}{2\sigma^2}\right)
\]

\noindent \textbf{Strategy 3: Semantic CLIP-based Retrieval.} We embed both text and images into a shared multimodal space using CLIP-based encoders~\cite{Radford2021LearningTV}. Cosine similarity is used to retrieve image-text pairs that are semantically aligned with the query image, capturing matches even when visual appearance differs but descriptive content is similar.

\subsection{Candidate Aggregation and Filtering}

We merge the retrieval results of all three strategies into unified candidate pools. Specifically, we define three existence-based multisets:
\[
\mathcal{M}_A = \{ l_i^{\text{raw}} \}, \quad
\mathcal{M}_B = \{ l_j^{\text{edge}} \}, \quad
\mathcal{M}_C = \{ l_m^{\text{clip}} \}
\]
We then perform deduplication and sorting on the union of the three sets:
\[
T = \text{sort}_{\leq}(\mathcal{M}_A \uplus \mathcal{M}_B \uplus \mathcal{M}_C) = (t_1, t_2, \dots, t_n)
\]
Here, $n = |\mathcal{M}_A \uplus \mathcal{M}_B \uplus \mathcal{M}_C|$ is the number of unique elements, and the elements are ordered by dictionary or numerical order. Finally, we truncate the list to retain only the top $m$ elements.



\begin{figure*}[t]
 \centering
 \vspace{-10mm}
 \includegraphics[width = \linewidth]{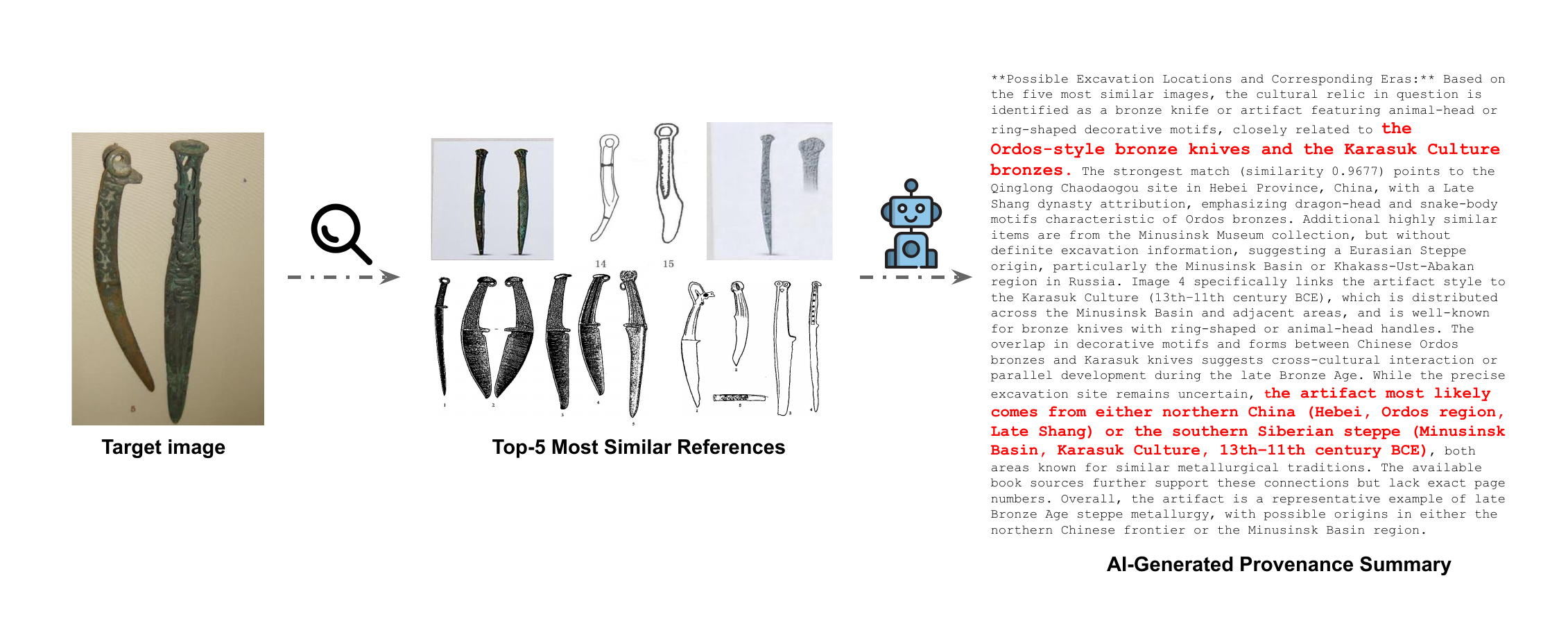}
    \vspace{-13mm}
    \caption{Example of system output for a Bronze Age bronze artifact. Left: input target image. Center: top-5 most similar reference objects retrieved from the external database, showing stylistic correspondences to Ordos-style bronzes and Karasuk Culture knives. Right: AI-generated provenance summary synthesizing visual matches and reference data to infer likely excavation regions (northern China or southern Siberia), estimated chronology (Late Shang to 13th–11th century BCE), and cultural affiliations.}
    \label{fig:result_demo}
\end{figure*}

\subsection{Inference and Reasoning}
The filtered candidates $M \subseteq T_{\text{top} \, m}$ are passed into the VLM for structured reasoning and synthetis. 

\noindent \textbf{Phase 1: Per-Candidate Interpretation.} The VLM analyzes each candidate individually, integrating visual features and textual context to infer key attributes: likely excavation site, estimated cultural period, similarity rationale, and bibliographic reference (including page number). Outputs are structured as metadata summaries in JSON file. This step significantly reduces the risk of hallucinations and alleviates token length constraints~\cite{belem-etal-2025-single}.

\noindent \textbf{Phase 2: Cross-Candidate Reasoning and Judgment.} In the second phase, the target image is fed into the model along with all Phase 1 outputs. The model determines the most likely excavation site and probable historical period for the object, drawing from correlations across both visual similarity and textual evidence extracted from the references. Furthermore, the VLM identifies the most relevant reference source that supports its inference, specifying both the document and the exact page number where a similar artifact or description is found. Finally, the model provides an interpretive justification for its prediction, highlighting key visual and textual cues that underpin its reasoning. 

\section{Evaluation}
\label{sec:eval}

\subsection{Experimental Settings}
\noindent \textbf{Datasets, Models, and Tasks.} We choose GPT-4o~\cite{Hurst2024GPT4oSC} as the VLM in the system. The reference material comprises eight exhibition catalogues and scholarly manuscripts. These publications, produced in Chinese, English, Russian, and French, constitute standard reference works routinely employed in professional training programs and advanced research contexts. Each catalogue provides systematic documentation through photographic plates and technical line drawings, accompanied by detailed chronological assessments and provenance data for individual object types.

\noindent \textbf{Expert Evaluation Criteria.} Five domain experts were recruited to evaluate the AI-generated outputs for the first thirty objects in the dataset. The evaluation framework is structured around two primary research questions:
\textbf{Q1}, assessing the identification of stylistically similar reference objects; and
\textbf{Q2}, assessing the generation of chronological, geographical, and archaeological cultural attributions.
The evaluation employed a structured four-level scoring scheme, which Score 4 represents highly meaningful and Score 1 suggests not meaningful.

\begin{figure}[ht]
    \centering
    \includegraphics[width=\linewidth]{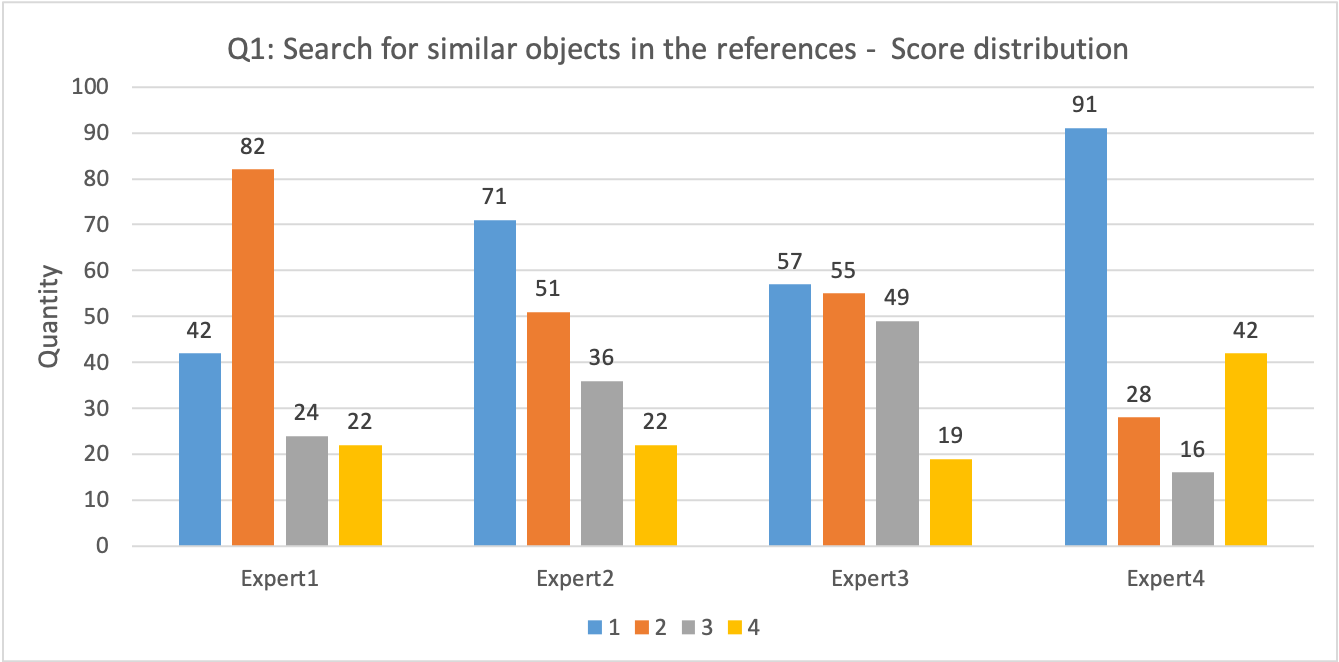}
    \vspace{0.3cm} 
    \includegraphics[width=\linewidth]{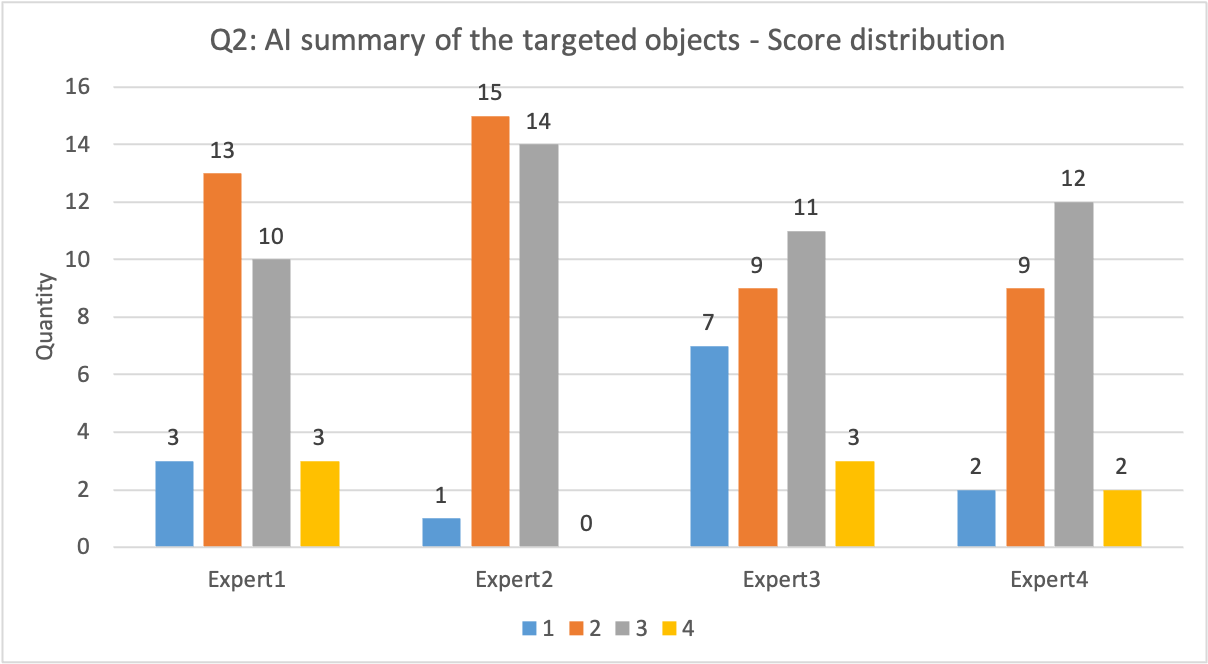}
    \caption{Expert evaluation score distributions for Q1 and Q2.}
    \vspace{-7mm}
    \label{fig:q1_q2_scores}
\end{figure}

\subsection{Evaluation Results}
Figure~\ref{fig:result_demo} provides a demonstration of system output for a target image. As Figure~\ref{fig:q1_q2_scores} shows, experts reckon that approximately 63\% of the retrieved images achieved meaningful results (Scores 2-4). However, the proportions of images receiving Score 3 (17.7\%) and Score 4 (14.9\%) remain relatively low compared to Score 2 (30.6\%).

The performance differential between Q1 and Q2 outcomes presents a particularly noteworthy finding. While visual similarity identification exhibited the aforementioned limitations, the system's capacity for generating chronological and geographical conclusions demonstrated markedly superior performance. The proportion of outcomes receiving the lowest rating decreased to approximately 10\% for Q2 evaluations, while nearly 46\% of the AI-generated attributions achieved Score 3 or higher. This disparity indicates the algorithm excels at synthesizing typological information into scholarly conclusions, suggesting greater utility for supporting expert decisions than replacing traditional comparative analysis.

\section{Conclusion}
We present a RAG-based system to assist provenance analysis of archaeological artifacts. Expert evaluation shows that while visual retrieval yields mixed performance, the system achieves notably stronger results in generating chronological and cultural attributions, providing researchers with interpretable and actionable outputs. Future work will explore integrating expert evaluation policies directly into the system to further enhance reliability and domain alignment.

{
    \small
    \bibliographystyle{ieeenat_fullname}
    \bibliography{main}
}

\end{document}